\documentclass[aps,prl,superscriptaddress,twocolumn]{revtex4-1}

\usepackage[colorlinks=true,linkcolor=blue,urlcolor=blue,citecolor=blue]{hyperref} 

\usepackage{amsmath}
\usepackage[pdftex]{graphicx}
\usepackage{color}

\begin{document}

\title{Resonantly phase-matched Josephson junction traveling wave parametric amplifier}

\author{Kevin O'Brien}
\affiliation{Nanoscale Science and Engineering Center, University of California, Berkeley, California 94720, USA}
\author{Chris Macklin}
\affiliation{Quantum Nanoelectronics Laboratory, Department of Physics, University of California, Berkeley, California 94720, USA}
\author{Irfan Siddiqi}
\affiliation{Quantum Nanoelectronics Laboratory, Department of Physics, University of California, Berkeley, California 94720, USA}
\author{Xiang Zhang}
\email{xiang@berkeley.edu}
\affiliation{Nanoscale Science and Engineering Center, University of California, Berkeley, California 94720, USA}
\affiliation{Materials Sciences Division, Lawrence Berkeley National Laboratory, Berkeley, California 94720, USA}

\begin{abstract}  We develop a technique to overcome phase-mismatch in Josephson-junction traveling wave parametric amplifiers in order to achieve high gain over a broad bandwidth. Using ``resonant phase matching," we design a compact superconducting device consisting of a transmission line with sub-wavelength resonant inclusions that simultaneously achieves a gain of 20 dB, an instantaneous bandwidth of 3 GHz, and a saturation power of -98 dBm. Such an amplifier is well-suited to cryogenic broadband microwave measurements such as the multiplexed readout of quantum coherent circuits based on superconducting, semiconducting, or nano-mechanical elements as well as traditional  astronomical detectors.  
\end{abstract}

\maketitle

Josephson parametric amplifiers\cite{castellanos-beltran_widely_2007,
bergeal_phase-preserving_2010,hatridge_dispersive_2011,roch_widely_2012,eichler_quantum_2014} routinely approach quantum-noise-limited performance \cite{louisell_quantum_1961,levenson-falk_dispersive_2013,castellanos-beltran_amplification_2008,mallet_quantum_2011}, and are currently used in sensitive experiments requiring high-fidelity detection of single-photon-level microwave signals, such as the readout and feedback control of superconducting quantum bits \cite{murch_observing_2013,vijay_stabilizing_2012,johnson_heralded_2012,slichter_jumps_2011,hatridge_quantum_2013,campagne-ibarcq_persistent_2013,riste_determinisitc_2013,riste_initialization_2012}, and magnetometry with the promise of single-spin resolution \cite{hatridge_dispersive_2011}. To obtain a large parametric gain, the interaction time with the material nonlinearity\textemdash the order-unity nonlinear inductance of the Josephson junction\textemdash must be maximized. Current Josephson parametric devices increase the interaction time by coupling the junction to a resonant cavity albeit at the expense of instantaneous bandwidth. In contrast, traveling wave parametric amplifiers\cite{cullen_travelling-wave_1958,cullen_theory_1959,sorenssen_theoretical_1962,tien_parametric_1958} (TWPAs) achieve long interaction times by utilizing long propagation lengths rather than employing multiple bounces in a cavity, thereby avoiding the inherent gain-bandwidth tradeoff associated with cavity based devices. A major challenge in the design of TWPAs, however, is that optimum parametric gain is only achieved when the amplification process is phase matched. TWPAs based on Josephson junctions have been investigated theoretically\cite{feldman_parametric_1975,sweeny_travelling-wave_1985,yaakobi_parametric_2013,yaakobi_erratum_2013} and experimentally\cite{wahlsten_parametric_1977,yurke_lownoise_1996,macklin_josephson_2014}, but have not demonstrated sufficient gain, in part due to phase-matching limitations,  to replace existing semiconductor amplifier technology. TWPAs based on the weaker nonlinear kinetic inductance of thin titanium nitride wires and phase matched through periodic loading have also been demonstrated\cite{ho_eom_wideband_2012,bockstiegel_development_2013}, but require significantly longer propagation lengths and higher pump powers to achieve comparable gain. In this paper, we show that by adding a resonant element into the transmission line, phase matching and exponential gain can be achieved over a broad bandwidth.

\begin{figure}
\includegraphics[]{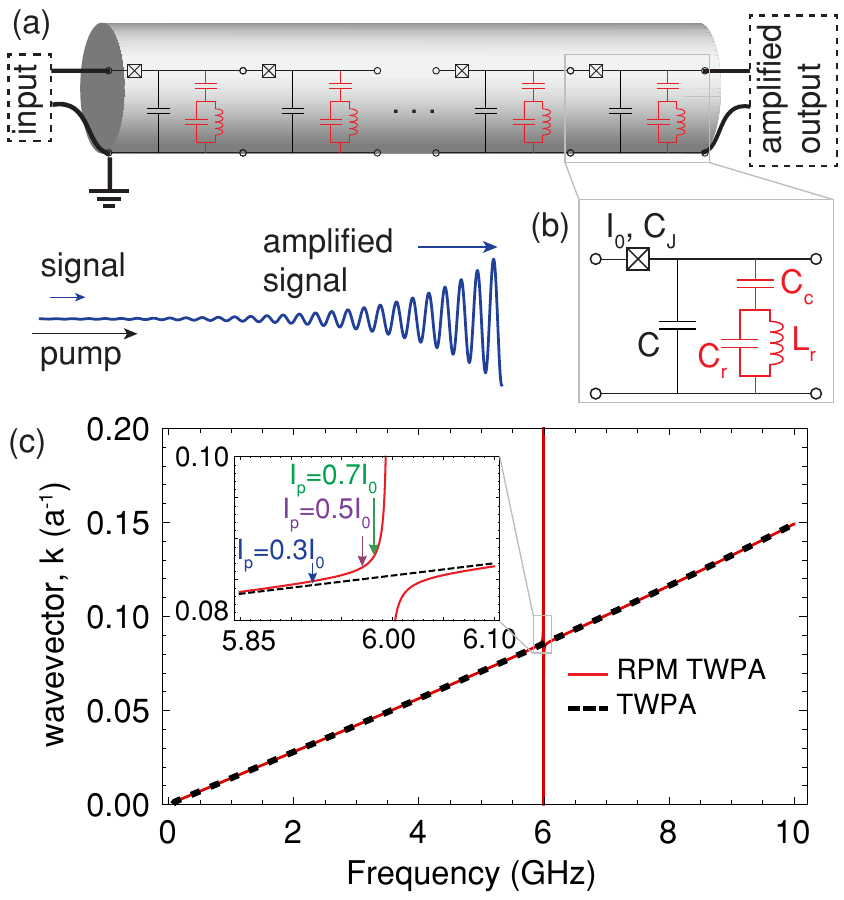}
\caption{Resonantly phase-matched traveling wave parametric amplifier (a) Signal photons are amplified through a nonlinear interaction with a strong pump as they propagate along the 2000 unit cell transmission line with a lattice period of $a$=10$\mathrm{\mu m}$. (b) In each unit cell a Josephson junction, a nonlinear inductor, is capacitively coupled to an $LC$ resonator. The circuit parameters are: $C_j$=329 fF, $L$=100 pH, $C$=39 fF, $C_c$=10 fF, $C_r$=7.036 pF, $L_r$=100 pH, $I_0$=3.29 $\mathrm{\mu A}$ (c) The $LC$ circuit opens a stop band (red) in the dispersion relation of the TWPA (black dashed) whose frequency depends on the circuit parameters. In the inset, we plot the pump frequency to phase-match a pump current of 0.3$I_0$ (blue), 0.5$I_0$ (purple), and 0.7$I_0$ (green), where $I_0$ is the junction critical current}
\label{fig:1}
\end{figure}




The proposed traveling wave parametric amplifier consists of a Josephson-junction-loaded transmission line (Fig.~\ref{fig:1}a) with a capacitively-coupled parallel LC resonator shunt to allow phase matching. The LC resonator shunt (colored red in Fig.~\ref{fig:1}a,b) creates a stop band (Fig.~\ref{fig:1}c, red) in the otherwise approximately linear dispersion relation (Fig.~\ref{fig:1}c, black dashed). In the presence of a strong co-propagating pump wave, a weak signal propagating in the TWPA is amplified through a four wave mixing interaction. Four wave mixing in the weak pump limit is perfectly phase-matched for a linear dispersion; however, a strong pump modifies the phase velocities through self and cross phase modulation, generating a phase mismatch and preventing exponential gain. We compensate this phase mismatch by tuning the pump frequency near the pole of the LC resonator. In a dissipationless system such as a superconducting circuit, a resonant element opens a stop band (inset of Fig.~\ref{fig:1}c), in which the wave vector is purely imaginary, surrounded by regions in which the wave vector is purely real and varies from 0 to $\pi/a$ where $a$ is the size of the unit cell. The wavevector of the pump can be set to arbitrary values by varying the frequency with respect to the resonance in order to eliminate the phase mismatch.

We now calculate the value of the phase mismatch and the expected device performance when phase matching is achieved. We use a first principles model for the nonlinear dynamics in the Josephson junction transmission line\cite{yaakobi_parametric_2013,yaakobi_erratum_2013} which has been validated by experiments\cite{macklin_josephson_2014}. By making the ansatz that the solutions are traveling waves, taking the slowly varying envelope approximation, and neglecting pump depletion, we obtain a set of coupled wave equations which describe the energy exchange between the pump, signal, and idler in the undepleted pump approximation (with the derivation in Eqs. \ref{eq:a1}-\ref{eq:a22} in the Appendix):
\begin{align}
\frac{{\partial {a_s}}}{{\partial x}} - i{\kappa _s}a_i^*{e^{i(\Delta {k_L} + 2{\alpha _p} - {\alpha _s} - {\alpha _i})x}} &= 0 \label{eq:1} \\
\frac{{\partial {a_i}}}{{\partial x}} - i{\kappa _i}a_s^*{e^{i(\Delta {k_L} + 2{\alpha _p} - {\alpha _s} - {\alpha _i})x}} &= 0 \label{eq:2}
\end{align}
where $a_s$ and $a_i$ are the signal and idler amplitudes, $\Delta {k_L} = 2{k_p} - {k_s} - {k_i}$ is the phase mismatch in the low pump power limit, and the coupling factors $\alpha_p$, $\alpha_s$, and $\alpha_i$ represent the change in the wave vector of the pump, signal, and idler due to self and cross phase modulation induced by the pump. The coupling factors depend on the circuit parameters (see Eqs.~\ref{eq:a14}, \ref{eq:a15}, and \ref{eq:a16} in the Appendix) and scale quadratically with the pump current. Maximum parametric gain is achieved when the exponential terms are constant: the phase mismatch, $\Delta k = \Delta {k_L} + 2{\alpha _p} - {\alpha _s} - {\alpha _i}$, must then be zero. The coupled wave equations (\ref{eq:1}), (\ref{eq:2}) are similar to the coupled amplitude equations for an optical parametric amplifier\cite{armstrong_interactions_1962} and have the solution:
\begin{equation}
{a_s}(x) = {a_s}(0)\left( {\cosh gx - \frac{{i\Delta k}}{{2g}}\sinh gx} \right){e^{i\Delta kx/2}} \label{eq:5}
\end{equation}
with the gain coefficient $g=\sqrt{\kappa_s \kappa^*_i -(\Delta k/2)^2}$. For zero initial idler amplitude and perfect phase matching, this leads to exponential gain, $a_s(x) \approx a_s(0)e^{gx}/2$. For poor phase matching $g$ is imaginary and the gain scales quadratically with length rather than exponentially. 

Without resonant phase matching, the parametric amplification is phase matched at zero pump power, but rapidly loses phase matching as the pump power increases. Neglecting dispersion and frequency dependent impedances, the exact expression for the phase mismatch can be simplified to yield $\Delta k \approx 2k_p-k_s-k_i - 2{k_p}\kappa$, where $\kappa  = \frac{a^2 k_p^2 \left| Z_{char} \right|^2}{16 L^2 \omega _p^2} \left( \frac{I_p}{I_0} \right)^2 $. The nonlinear process creates a pump power dependent phase mismatch which can be compensated by increasing the pump wave vector. 

\begin{figure}
\includegraphics[]{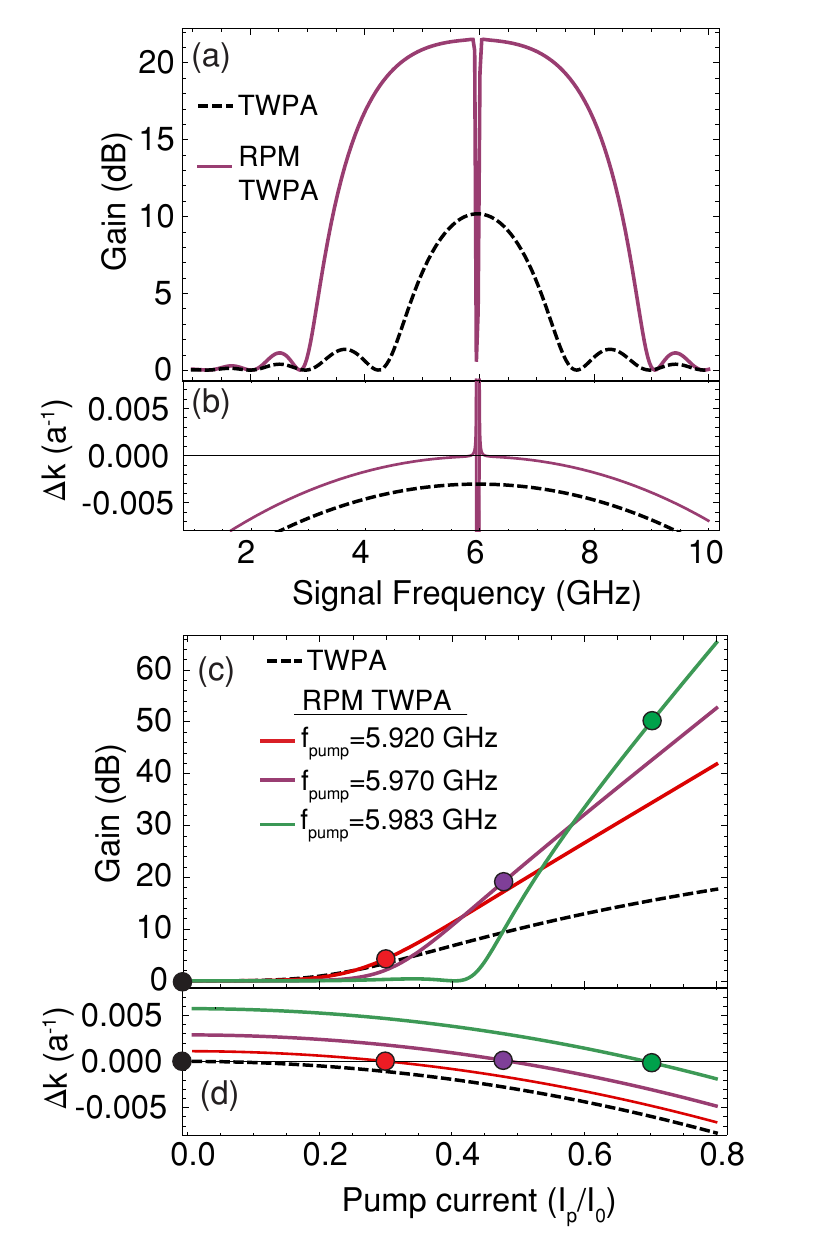}
\caption{Gain of the resonantly phase matched traveling wave parametric amplifier (RPM TWPA). (a) The gain as a function of signal frequency in dB with RPM (purple) and without (black dashed) for a pump current of 0.5$I_0$ and a pump frequency of 5.97 GHz. (b) The phase mismatch with (purple) and without (black dashed) RPM. (c) The peak gain as a function of pump current without RPM (black dashed) and with RPM for three different pump frequencies, which phase match the parametric amplification for pump currents of 0.3 $I_0$ (red), 0.5 $I_0$ (purple), and 0.7 $I_0$ (green). (d) The phase mismatch as a function of pump current. The dots mark the pump current where the parametric amplification is perfectly phase matched.}
\label{fig:2}
\end{figure}

In Fig.~\ref{fig:2}, we show the increase in gain due to resonant phase matching for the device described in Fig~\ref{fig:1}. Resonant phase matching increases the gain by more than one order of magnitude from 10 dB to 21 dB (Fig.~\ref{fig:2}a) for a pump current of half the junction critical current and a pump frequency, 5.97 GHz, on the lower frequency tail of the resonance as shown in the inset of Fig.~\ref{fig:1}c. The increase in the pump wave vector due to the resonance compensates the phase mismatch from cross and self phase modulation (Fig.~\ref{fig:2}b, black dashed) leading to perfect phase matching near the pump frequency (Fig.~\ref{fig:2}b, purple). For higher pump currents, the benefits are even more pronounced: the RPM TWPA achieves 50 dB of gain (compared to 15 dB for the TWPA) with a pump current of 0.7$I_0$ (Fig.~\ref{fig:2}c). Achieving 50 dB of gain over a 3 GHz bandwidth would require a larger junction critical current than used here to prevent gain saturation by vacuum photons. By varying the pump frequency relative to the resonance, the parametric amplification can be phase matched for arbitrary pump currents (Fig.~\ref{fig:2}d). Due to this ability to tune the pump phase mismatch over the entire range of possible wavevectors, this technique is highly flexible and can accommodate a variety of pumping conditions.

\begin{figure}
\includegraphics[]{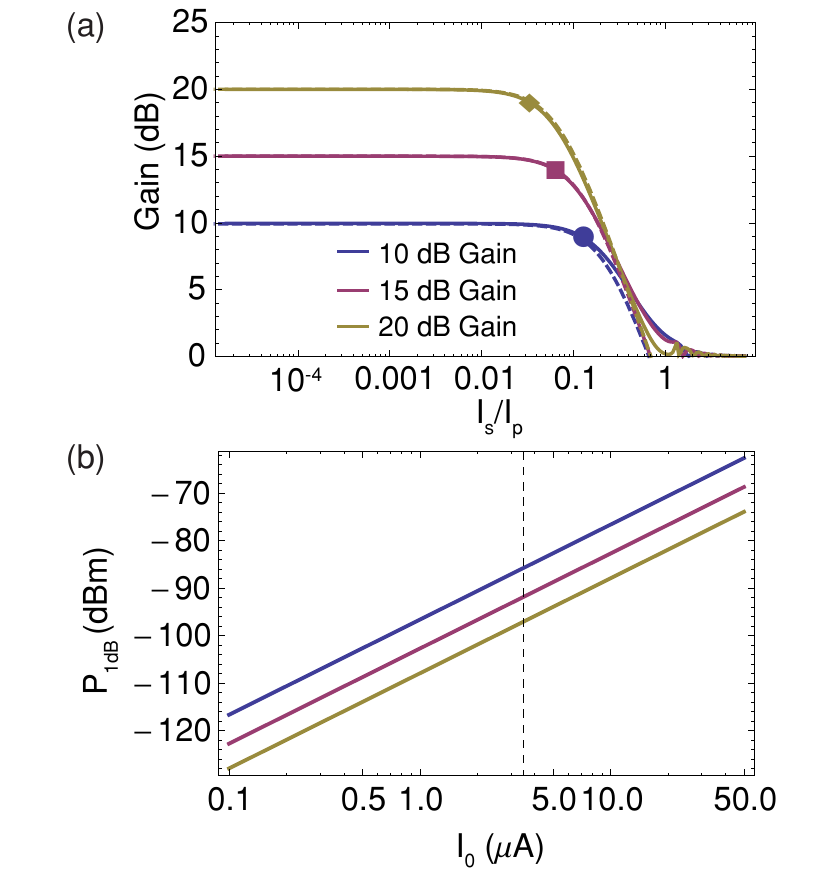}
\caption{Effect of pump depletion on dynamic range. (a) The gain as a function of input signal current (normalized to the pump current) for a small signal gain of 10, 15, and 20 dB obtained with a pump current of $0.5 I_0$ and device lengths of 1150, 1530, and 1900 unit cells. The approximation for the gain depletion (dashed lines) from Eq.~\ref{eq:7} is in excellent agreement with the result obtained by solving the full nonlinear dynamics (solid lines). (b) $P_{1dB}$, the input signal power where the gain decreases by 1 dB, as a function of junction critical current $I_0$ with the pump current fixed at $0.5 I_0$. The black dashed line corresponds to the device considered in this article which has a gain compression point of $P_{1dB}$=-87,-93, and -98 dBm for a gain of 10, 15, and 20 dB for a junction critical current of $I_0=3.29~\mathrm{\mu A}$.}
\label{fig:3}
\end{figure}

We now examine the scaling relations for the gain in order to obtain the optimum gain through engineering the linear and nonlinear properties of the transmission line. Simplifying the expression for the gain by assuming perfect phase matching and neglecting the effects of the resonant element and the junction resonance on the dispersion we find that the exponential gain coefficient is directly proportional to the wave vector $g\propto k_p I_p^2/I_0^2$. Full expressions for the wave vector and characteristic impedance are given by Eq.~\ref{eq:k} and Eq.~\ref{eq:zchar}. Thus, for a fixed pump strength relative to the junction critical current, the gain coefficient is proportional to the electrical length. In other words, a larger wave vector and thus slower light leads to a larger effective nonlinearity due to the higher energy density; this effect is well known in photonic crystals\cite{soljacic_photonic-crystal_2002}. For convenient integration with commercial electronics the characteristic impedance is designed to be $Z_{char} \approx \sqrt{L/(C+C_c)} \approx 50 \Omega$ which fixes the ratio of the inductance and capacitance. The wave vector is proportional to the product of the capacitance and inductance $k \approx \omega/a \sqrt{L(C+C_c)}$. Increasing both the capacitance and inductance or decreasing the unit cell size are effective strategies for increasing the gain per unit length while maintaining impedance matching for a 50 ohm load. The capacitors and inductors take a finite amount of space which constrains the minimum size of the unit cell. The current design represents a trade off between unit cell size and component values which is convenient to fabricate. 

Next we consider the dynamic range of the amplifier. The upper limit of the dynamic range of a parametric amplifier is given by pump depletion: the pump transfers energy to the signal and idler which reduces the parametric gain. To investigate this regime, we solve for the coupled wave equations without the undepleted pump approximation, resulting in four coupled nonlinear differential equations (Eqs.~\ref{eq:a4:2}-\ref{eq:a4:5} in the Appendix). We solve these equations by transforming them to real differential equations and expressing them as a Jacobi elliptic integral\cite{chen_four-wave_1989}. The gain as a function of input signal power calculated from Eqs. \ref{eq:a4:2}-\ref{eq:a4:5} (solid lines in Fig.~\ref{fig:3}) is in excellent agreement with the approximate yet general solution for pump depletion (dashed lines in Fig.~\ref{fig:3}) in a four photon parametric amplifier\cite{kylemark_semi-analytic_2006}:
\begin{equation}
G=\frac{G_0}{1+2 G_0 I_{s}^2/I_{p}^2} \label{eq:7}
\end{equation}
where $G_0$ is the small signal gain in linear units and $I_{s}$ and $I_{p}$ are the input signal and pump currents. From Eq.~\ref{eq:7}, the gain compression point is approximately $P_{1dB}=P_{p}/(2 G_0)$. Thus, the threshold for gain saturation is independent of the specific device configuration and depends only on the small signal gain and the pump power. For the considered device, the gain as a function of input signal current is plotted for three values of the small signal gain in Fig.~\ref{fig:3}a. The signal current at which the gain drops by 1 dB is marked on the curves of Fig.~\ref{fig:3}a. For the considered device with a pump current of $1.75\mu A$ the signal power where the gain decreases by 1 dB is $-87$, $-93$, and $-98$ dBm for a small signal gain of 10, 15, and 20 dB, respectively. These gain compression points are consistent with the approximate relation with the pump power of -69 dBm. The dynamic range of the TWPA is significantly higher than a cavity based Josephson parametric amplifier with the same junction critical current since the lack of a cavity enables a higher pump current before the Josephson junction is saturated.

To further increase the threshold for gain saturation, the junction critical current can be scaled up, as seen in Fig.~\ref{fig:3}b. However, increasing the junction critical current decreases the inductance which reduces the wave vector; leading to a weaker nonlinearity. One potential solution is to use N Josephson junctions each with a critical current $I_0/N$ in series as a superinductor\cite{masluk_microwave_2012}. The 1 dB power then scales as $\sqrt{N}$ leading to a larger dynamic range at the expense of more complex fabrication. 

We now estimate the potential for pump distortion due to third harmonic or higher harmonic generation. This was suggested as a mechanism leading to self-steepening and instabilities in parametric amplification\cite{landauer_parametric_1960,landauer_shock_1960}. We find that the third harmonic is poorly phased matched and less than $0.1\%$ of the pump is converted to the third harmonic. Due to the junction resonance at $1/\sqrt {L C_J}  \approx 2\pi \cdot \mathrm{27.7~GHz}$ , the dispersion  contains a significant quadratic component which increases the phase mismatch for third harmonic generation (Fig~\ref{fig:th}a,b in the Appendix), and prevents the formation of higher harmonics due to the stop band above the junction resonance. We calculate the third harmonic amplitude to be ${a_{th}} = (1 - {e^{ix\Delta k}}){\kappa _2}/\Delta k$ where $\Delta k = \Delta {k_L} + 3{\kappa _0} - {\kappa _1}$ and $\Delta {k_L} = 3{k_p} - {k_{th}}$.  The couplings are defined in Eq.~\ref{eq:th:7}-\ref{eq:th:10} and are different from the couplings for parametric amplification. The third harmonic amplitude oscillates with device length and pump power (Fig~\ref{fig:th}c,d in the Appendix) and has a maximum amplitude given by coupling ${\kappa _2}/\Delta k$. For a pump current of one half the junction critical current, the third harmonic power is 3 orders of magnitude weaker than the pump power. As seen in the above analysis, the third harmonic is too weak to cause significant pump depletion for the system under consideration.


\begin{figure}
\includegraphics[]{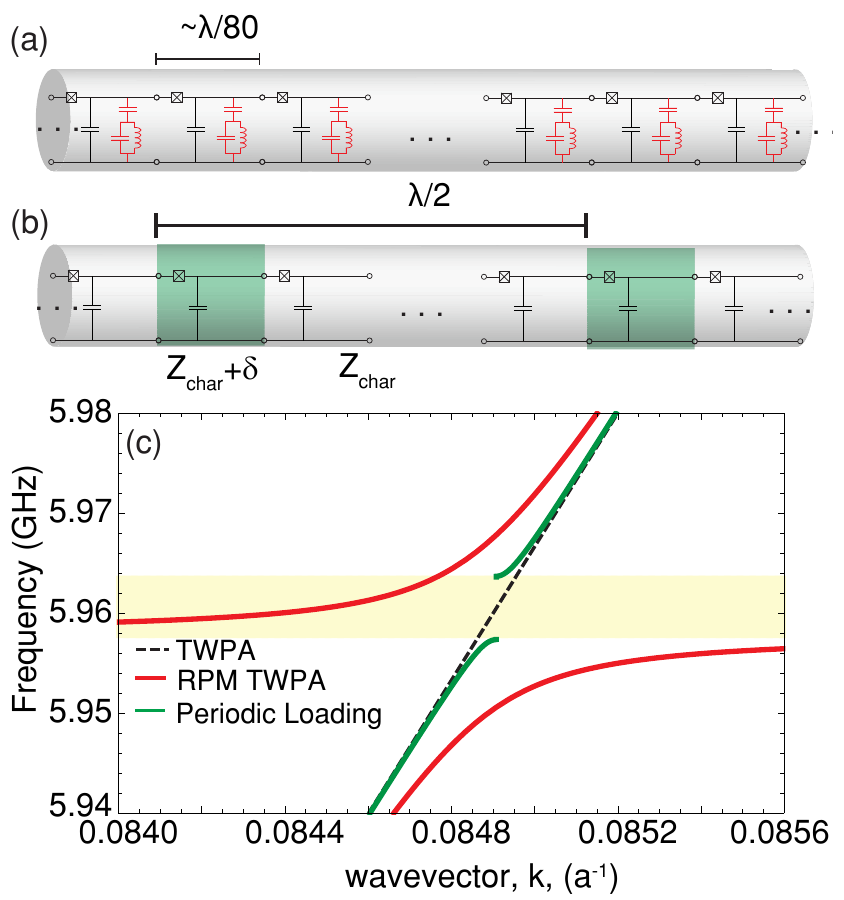}
\caption{Band structure of resonantly phase matched and periodically loaded TWPAs. The parametric amplifier without a (a, black only) resonant element, (a, black and red) with a resonant element, and (b) with periodic loading:  every 37th unit cell has a slightly different capacitance and inductance (green). (c) The wave vector as a function of frequency for the TWPA (black dashed), RPM TWPA (red), and TWPA with periodic loading (green). The yellow shaded region indicates the photonic band gap due to the periodic loading.  The main difference between the photonic band gap and the resonator is the edge of the Brillouin zone. For the resonant element, the zone boundary is at $\pi/a$ while the other periodically loaded transmission line has zone boundary at $\pi/(37a)$ which is determined by the period of the loading. The effective momentum due to the periodic loading is close to $k_p$, which may phase match backward parametric amplification.}
\label{fig:4}
\end{figure}

Resonant phase matching has an important advantage over dispersion engineering through periodic loading which has been used to phase match TWPAs based on the weaker nonlinear kinetic inductance\cite{ho_eom_wideband_2012,bockstiegel_development_2013}. A disadvantage of periodic loading is the potential phase matching of backward parametric amplification. Dispersion engineering through periodic loading opens a photonic band gap (Fig.~\ref{fig:4}) near the pump frequency and through band bending changes the pump wave propagation constant to phase match forward parametric amplification. Periodic loading creates an effective momentum inversely proportional to the periodicity of the loading $G = 2\pi /\Lambda$ where $\Lambda$ is the periodicity of the loading and $G$  is the reciprocal lattice vector. The periodicity is chosen so that the stop band is at $G/2 \approx {k_p}$.  In such a periodic system the phase matching relation needs only to be satisfied up to an integer multiple of the reciprocal lattice vector\cite{bloembergen_nonlinear_1970}. As can be seen from the phase matching relation, the effective momentum from the lattice phase-matches the parametric amplification process for a backward propagating signal $\Delta {k_{L,b}} = 2{k_p} + {k_s} + {k_i} + nG \approx 0$ for $n=-2$. Under this condition, any backward propagating photons present in the system will be amplified, leading to gain ripples and a reduced  threshold for parametric oscillations. Due to imperfect impedance matching over the operating band of the amplifier, a weak standing wave condition will be set up in the nonlinear transmission line due to the return loss at the output and input.  If the return loss in dB at the transition from nonlinear line to linear line is $R$, then the magnitude of the standing signal will be of order $2R$.  However, the signal experiences some gain (in dB) in the forward and reverse directions, $G_f$ and $G_r$.  If $G_f + G_r + 2R$ approaches unity, the device becomes a parametric oscillator. The proposed resonant phase matching technique phase matches only the forward parametric amplification process, so the maximum gain before the onset of parametric oscillations may be higher than in a device utilizing periodic loading.

In conclusion, we have developed a traveling wave parametric amplifier which is phase matched by sub-wavelength resonant elements and achieves 20 dB of gain, 3 GHz of bandwidth, and a saturation power ($P_{1dB}$) of -98 dBm. This device is well suited to multiplexed readout of quantum bits and astronomical detectors. Applying metamaterial design techniques to nonlinear superconducting systems may yield a number of useful devices for circuit quantum electrodynamics such as backward parametric amplifiers or mirror-less optical parametric oscillators\cite{popov_microscopic_2009}.

\begin{acknowledgments}
\section{Acknowledgments}
The authors wish to acknowledge L. Friedland and O. Yaakobi for prior theory work and N. Antler for useful discussions. This research was supported in part by a Multidisciplinary University Research Initiative from the Air Force Office of Scientiﬁc Research (AFOSR MURI Award No. FA9550-12-1-0488),  the Army Research Office (ARO) under grant W911NF-14-​1-0078 and the Office of the Director of National Intelligence (ODNI), Intelligence Advanced Research Projects Activity (IARPA), through the Army Research Office. All statements of fact, opinion or conclusions contained herein are those of the authors and should not be construed as representing the official views or policies of IARPA, the ODNI or the US government.
\end{acknowledgments}

\bibliography{twpa}

\section{Appendix 1: Parametric amplification}

Here we derive the coupled wave equations for a traveling wave parametric amplifier. The nonlinear wave equation is (Eq.~22 from Ref. \cite{yaakobi_parametric_2013}):
\begin{equation}
{C_0}\frac{{{\partial ^2}\phi }}{{\partial {t^2}}} - \frac{{{a^2}}}{L}\frac{{{\partial ^2}\phi }}{{\partial {x^2}}} - {C_j}{a^2}\frac{{{\partial ^4}\phi }}{{\partial {x^2}\partial {t^2}}} = \frac{{{a^4}}}{{2I_0^2{L^3}}}\frac{{{\partial ^2}\phi }}{{\partial {x^2}}}{\left( {\frac{{\partial \phi }}{{\partial x}}} \right)^2} \label{eq:a1}
\end{equation}
We take the ansatz that the solutions will be forward propagating waves of the form:
\begin{multline}
\phi  = \frac{1}{2} [ A_p(x)e^{i(k_p x + \omega _p t)} + A_s(x)e^{i(k_s x + \omega _s t)} +\\ A_i(x)e^{i(k_ix + \omega _it)} + c.c]
\end{multline}
Where $A_m$ is the slowly varying amplitude, $k_m$ is the wave vector, and $\omega_m$ is the angular frequency. We substitute the above expression into the nonlinear wave equation then make the following approximations: 
\begin{enumerate}
\item	Neglect the second derivatives of the slowly varying amplitudes using the slowly varying envelope approximation: $\left| {\frac{{{\partial ^2}{A_m}}}{{\partial {x^2}}}} \right| \ll \left| {{k_m}\frac{{\partial {A_m}}}{{\partial x}}} \right|$.
\item	Neglect the first derivatives of the slowly varying amplitudes on the right side of the nonlinear wave equation (ie, in the nonlinear polarizability): $\left| {\frac{{\partial {A_m}}}{{\partial x}}} \right| \ll \left| {{k_m}{A_m}} \right|$.  
\end{enumerate}
Considering only the left side of Eq.~\ref{eq:a1} and separating out the terms that oscillate at the pump, signal, and idler frequencies we get the following equation:
\begin{align}
\Biggl[ \frac{a^2 e^{i(t\omega_m + k_m x)} k_m^2}{2L} - 
\frac{1}{2} C_0 e^{i(t\omega _m + k_m x)}\omega _m^2 - \notag\\  \frac{1}{2} a^2 C_j e^{i(t\omega _m + k_m x)}k_m^2 \omega _m^2 \Biggr] A_m(x)  \notag \\
+ \Biggl[ i a^2 C_j e^{i(t\omega _m + k_m x)}k_m \omega _m^2 -
\frac{i a^2 e^{i(t\omega _m + k_m x)} k_m}{L}  \Biggr] \frac{\partial A_m(x)}{\partial x} \label{eq:a2}
\end{align}
where $m=p,s,i$. Defining the wave vector as ${k_m} = \frac{{{\omega _m}\sqrt {{C_0}L} }}{{a\sqrt {1 - {C_j}L{\omega _m}} }}$, Eq.~\ref{eq:a2} simplifies to:
\begin{equation}
 - \frac{{i{C_0}{\omega _m}^2}}{{{k_m}}}{{\rm{e}}^{{\rm{i}}(t{\omega _m} + {k_m}x)}} \frac{{\partial {A_m}(x)}}{{\partial x}} \label{eq:a3}
\end{equation}
Now we consider the nonlinear component (the right side of Eq.~\ref{eq:a1}). The propagation equation for the pump is:
\begin{equation}
\frac{{\partial {A_p}}}{{\partial x}} - \frac{{i{a^4}{k_p}^5}}{{16{C_0}{I_0}^2{L^3}\omega _p^2}}{A_p}^2A_p^* = 0 \label{eq:a4}
\end{equation}
where we have neglected the terms proportional to the amplitudes of the signal and idler as they are much smaller than the pump field. The propagation equation for the signal and idler, neglecting terms which are quadratic in the signal and idler amplitudes:
\begin{align}
\frac{{\partial {A_s}}}{{\partial x}} - i\frac{{{a^4}{k_p}^2{k_s}^3}}{{8{C_0}{I_0}^2{L^3}{\omega _s}^2}}{A_p}A_p^*{A_s} -  \notag\\ i\frac{{{a^4}{k_p}^2(2{k_p} - {k_i}){k_s}{k_i}}}{{16{C_0}{I_0}^2{L^3}{\omega _s}^2}}{A_p}^2A_i^*{{\rm{e}}^{{\rm{i}}\Delta {k_L}x}} = 0 \label{eq:a5}\\
\frac{{\partial {A_i}}}{{\partial x}}  - i\frac{{{a^4}{k_p}^2{k_i}^3}}{{8{C_0}{I_0}^2{L^3}{\omega _i}^2}}{A_p}A_p^*{A_i} - \notag\\ i\frac{{{a^4}k_p^2(2{k_p} - {k_s}){k_s}{k_i}}}{{16{C_0}{I_0}^2{L^3}{\omega _i}^2}}{A_p}^2A_s^*{{\rm{e}}^{{\rm{i}}\Delta {k_L}x}} = 0 \label{eq:a6}
\end{align}
Now we solve for the pump propagation, assuming no loss, and obtain:
\begin{equation}
A_p(x) = A_{p,0}e^{i\frac{a^4 k_p^5 A_p A_p^*}{16 C_0 I_0^2 L^3\omega _p^2}x} \label{eq:a7}
\end{equation}
We substitute the solution for the pump field (Eq.~\ref{eq:a7}) into Eqs.~\ref{eq:a5} and \ref{eq:a6}:
\begin{align}
A_p(x) = A_{p,0}e^{i\alpha _p x} \label{eq:a8}\\
\frac{\partial A_s}{\partial x} - i\alpha _s A_s - i\kappa _s A_i^* e^{i(\Delta k_L + 2\alpha _p)x} = 0 \label{eq:a9}\\
\frac{\partial A_i}{\partial x} - i\alpha _i A_i - i\kappa _i A_s^* e^{i(\Delta k_L + 2\alpha _p)x} = 0 \label{eq:a10}
\end{align}
where the couplings are defined as:
\begin{align}
\alpha_s = \frac{2\kappa k_s^3 a^2}{LC_0\omega _s^2} && \kappa _s = \frac{\kappa (2 k_p - k_i)k_s k_i a^2}{LC_0\omega _s^2} \label{eq:a11}\\
\alpha_i = \frac{2\kappa k_i^3a^2}{LC_0\omega _i^2} && \kappa _i = \frac{\kappa (2k_p - k_s)k_s k_i a^2}{LC_0\omega _i^2} \label{eq:a12}\\
\alpha_p = \frac{\kappa k_p^3 a^2}{LC_0\omega _p^2} && \kappa  = \frac{a^2 k_p^2 A_{p,0}A_{p,0}^*}{16I_0^2 L^2} \label{eq:a13}
\end{align}
To generalize these equations for arbitrary circuits, we make the substitution ${C_0} = 1/(i\omega Z_2)$ and express the pump amplitude in terms of the characteristic impedance and pump current: $A_{p,0}=I_p Z_{char}/\omega_p$. The couplings are now:
\begin{align}
\alpha_s = \frac{2\kappa k_s^3 a^2 i Z_2(\omega_s)} {L\omega _s} && \kappa _s = \frac{\kappa (2 k_p - k_i)k_s k_i i Z_2(\omega_s) a^2}{L\omega _s} \label{eq:a14}\\
\alpha_i = \frac{2\kappa k_i^3a^2 i Z_2(\omega_i)} {L\omega _i} && \kappa _i = \frac{\kappa (2k_p - k_s)k_s k_i i Z_2(\omega_i) a^2}{L\omega _i} \label{eq:a15}\\
\alpha_p = \frac{\kappa k_p^3 a^2 i Z_2(\omega_p)} {L\omega _p} && \kappa  = \frac{a^2 k_p^2 |Z_{char}|^2}{16L^2 \omega_p^2} \left(\frac{I_p}{I_0}\right)^2 \label{eq:a16}
\end{align}

We solve the coupled amplitude equations (Eqs.~\ref{eq:a9} and \ref{eq:a10}) by making the substitutions $A_s=a_s e^{i\alpha_s x}$ and $A_i=a_i e^{i\alpha_i x}$ to obtain:
\begin{align}
\frac{{\partial {a_s}}}{{\partial x}} - i{\kappa _s}a_i^*{e^{i(\Delta {k_L} + 2{\alpha _p} - {\alpha _s} - {\alpha _i})x}} &= 0 \label{eq:a17} \\
\frac{{\partial {a_i}}}{{\partial x}} - i{\kappa _i}a_s^*{e^{i(\Delta {k_L} + 2{\alpha _p} - {\alpha _s} - {\alpha _i})x}} &= 0 \label{eq:a18}
\end{align}
These equations are analogous to the coupled amplitude equations for an optical parametric amplifier, which have the following solution\cite{armstrong_interactions_1962}:
\begin{align}
a_s(x) = \Biggl[ a_s(0)\left(\cosh gx - \frac{i\Delta k}{2g}\sinh gx \right) +\notag \\ \frac{i\kappa_s}{g}a_i^*(0)\sinh gx \Biggr] e^{i\Delta kx/2} \label{eq:a19}\\
a_i(x) = \Biggl[ a_i(0)\left(\cosh gx - \frac{i\Delta k}{2g}\sinh gx \right) + \notag \\ \frac{i\kappa_i}{g}a_s^*(0)\sinh gx \Biggr] e^{i\Delta kx/2} \label{eq:a20}
\end{align}
where $\Delta k$ and $g$ are defined as:
\begin{align}
\Delta k &= \Delta {k_L} + 2{\alpha _p} - {\alpha _s} - {\alpha _i} \notag \\
&= 2k_p-k_s-k_i + 2{\alpha _p} - {\alpha _s} - {\alpha _i}  \label{eq:a21}
\end{align}
\begin{equation}
g=\sqrt{\kappa_s \kappa^*_i -(\Delta k/2)^2}  \label{eq:a22}
\end{equation}

\section{Appendix 2: Linear properties}
For the considered circuit topology (Fig.~\ref{fig:1}b) the ABCD matrix is:
\[ \left( \begin{array}{cc}
1 & -Z_1 \\
-1/Z_2 & 1+Z_1/Z_2 \end{array} \right)\] 
where 
\begin{align}
Z_1=Z_L||Z_{C_j}=\left( \frac{1}{i\omega L}+i\omega C_j \right)^{-1}  \\
Z_2=Z_C||Z_{res} \\
Z_{res}=Z_{C_c}+Z_{C_r}||Z_{L_r} = \frac{1-(C_c+C_r)L_r\omega^2}{i\omega C_c(1-C_r L_r \omega^2)}
\end{align}
The wavevector and characteristic impedance in terms of the ABCD matrix elements are:
\begin{align}
k=\cos^{-1}{\frac{A+B}{2}}  \label{eq:k} \\
Z_{char}=\frac{(A-D)+\sqrt{(A+D+2)(A+D-2)}}{2C} \label{eq:zchar}
\end{align}

\section{Appendix 3: Third harmonic generation}
The general procedure applies to third harmonic generation as well. We make the undepleted pump approximation which means that we assume the third harmonic is always much weaker than the pump. Depending on the phase matching, this may not be the case in the experiment. In the undepleted pump approximation, the wave equations are:
\begin{align}
\frac{\partial A_p}{\partial x} - \frac{i a^4 k_p^5}{16 C_0 I_0 ^2 L^3 \omega _p^2} A_p^2 A_p^* = 0 \label{eq:th:1}\\
\frac{\partial A_{th}}{\partial x} - i\frac{a^4 k_p^2 k_{th}^3}{8 C_0 I_0^2 L^3 \omega _{th}^2} A_p A_p^* A_{th} + \notag \\
 i\frac{ a^4 k_p^4 k_{th}}{ 16 C_0 I_0^2 L^3 \omega _{th}^2} A_p^3 e^{i \Delta k_L x} = 0 \label{eq:th:2}
\end{align}
Using the same procedure of solving for the pump field
\begin{align}
A_p (x) = A_{p,0} e^{i\frac{a^4 k_p ^5 A_p A_p^*}{ 16 C_0  I_0 ^2L^3 \omega _p^2}x} \label{eq:th:3}\\
\frac{\partial A_{th}}{\partial x} - i\frac{a^4 k_p^2 k_{th}^3}{8 C_0 I_0^2 L^3 \omega _{th}^2} A_{p,0}A_{p,0}^* A_{th} \notag \\
+ i\frac{a^4 k_p^4 k_{th}}{16 C_0 I_0^2 L^3 \omega _{th}^2}A_{p,0}^3 e^{i\left(\Delta k_L + 3\frac{a^4 k_p ^5 A_p A_p^*}{16 C_0 I_0^2 L^3 \omega _p^2} \right)x} = 0 \label{eq:th:4}
\end{align}
and rewriting it in terms of the couplings
\begin{align}
{A_p}(x) &= {A_{p,0}}{e^{i{\kappa _0}x}} \label{eq:th:5}\\
\frac{{\partial {A_{th}}}}{{\partial x}} - i{\kappa _1}{A_{th}} + i{\kappa _2}{{\rm{e}}^{{\rm{i}}\left( {\Delta {k_L} + 3{\kappa _0}} \right)x}}&= 0 \label{eq:th:6}
\end{align}
with the couplings defined as:
\begin{align}
\kappa  = \frac{a^2 k_p^2 |Z_{char}|^2}{16 L^2 \omega _p^2} {\left( \frac{I_p}{I_0} \right)}^2 \label{eq:th:7} \\
\kappa _0 = \frac{\kappa k_p^3 i Z_2(\omega _p) a^2} {L \omega _p} \label{eq:th:8} \\
\kappa _1 = \frac{2\kappa k_{th}^3 i Z_2(\omega _{th})a^2 } {L \omega _{th}} \label{eq:th:9} \\
\kappa _2 = \frac{\kappa k_p^2 k_{th} i Z_2(\omega _{th}) a^2}{ L \omega _{th} } \left(\frac{|Z_{char}| I_p}{\omega _p I_0} \right) \label{eq:th:10}
\end{align}

\begin{figure}
\includegraphics[]{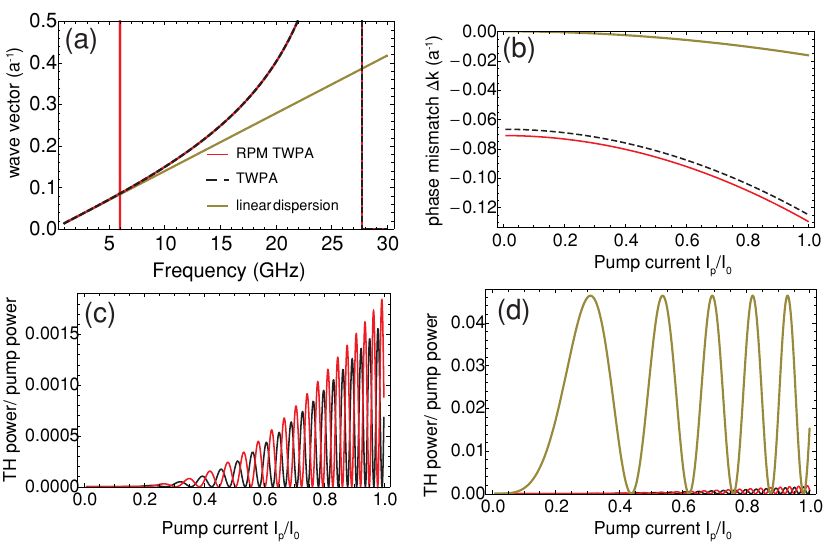}
\caption{Third harmonic. (a) The wave vector with the resonance (red), without the resonance (black dashed), and with a linearized dispersion (gold) equal to $\frac{\omega}{a} \sqrt {L(C + {C_c})} $  which neglects the junction resonance. (b) The phase mismatch corresponding to the dispersion relations in (a) for a pump of 5.97 GHz. As expected,  the linearized resonance is perfectly phase matched in the limit of zero pump power (gold). The more realistic dispersions with the resonance (red) and without (black dashed) have a large phase mismatch which gets larger as the pump power increases. (c) The relative third harmonic power increases with the pump power, but is negligible with (red) and without (black) the resonance.(d) The third harmonic power of the linearized resonance is significantly larger than in the realistic system.
}
\label{fig:th}
\end{figure}

We solve this making the substitutions  $a_{th} = A_{th} e^{i \kappa _1 x}$ and obtain:
\begin{equation}
\frac{{\partial {a_{th}}}}{{\partial x}} + i{\kappa _2}{{\rm{e}}^{{\rm{i}}\left( {\Delta {k_L} + 3{\kappa _0} - {\kappa _1}} \right)x}} = 0 \label{eq:th:11}
\end{equation}
which has the solution: 
\begin{align}
a_{th} &= \frac{(1 - e^{i x(\Delta k_L + 3\kappa _0 - \kappa _1)}) \kappa _2}{\Delta k_L + 3\kappa _0 - \kappa _1}\\ &= \frac{(1 - e^{i x\Delta k})\kappa _2}{\Delta k} \label{eq:th:12}
\end{align}
when the initial amplitude is zero , where $\Delta k = \Delta {k_L} + 3{\kappa _0} - {\kappa _1}$ and $\Delta k_L = 3 k_p - k_{th}$. 
In Fig.~\ref{fig:th} we plot the dispersion relation, phase mismatch, and the third harmonic power for the case with resonant phase matching, without, and with a fictitious linearized dispersion relation (equivalent to neglecting the junction resonance). The third harmonic is too weak to cause significant pump depletion for the system under consideration.

\section{Appendix 4: Parametric amplification including pump depletion}
We now consider the general solution including pump depletion effects and non-degenerate pumps. We now look for a traveling wave solution with two pump waves, a signal and an idler. 
\begin{align}
\phi  = \frac{1}{2} [ A_1(x)e^{i(k_1 x + \omega_1 t)} + A_2(x)e^{i(k_2 x + \omega _2 t)} + \notag \\ A_3(x)e^{i(k_3 x + \omega _3 t)} + A_4(x)e^{i(k_4 x + \omega _4 t)} + c.c] \label{eq:a4:1}
\end{align}
where $A_1$ and $A_2$ are the pumps and $A_3$ and $A_4$ are the idler and signal. Plugging this ansatz into the nonlinear wave equation (Eq.~\ref{eq:a1}) and making the slowly varying envelope approximation, we obtain the following coupled amplitude equations:
\begin{align}
\frac{{d{A_1}}}{{dx}} - i{\kappa _1}A_2^*{A_3}{A_4}{{\rm{e}}^{ - i\Delta kx}} - i{A_1}\sum\limits_{m = 1}^4 {{\alpha _{1m}}{A_m}A_m^*}  = 0 \label{eq:a4:2} \\
\frac{{d{A_2}}}{{dx}} - i{\kappa _2}A_1^*{A_3}{A_4}{{\rm{e}}^{ - i\Delta kx}} - i{A_2}\sum\limits_{m = 1}^4 {{\alpha _{2m}}{A_m}A_m^*}  = 0 \label{eq:a4:3} \\
\frac{{d{A_3}}}{{dx}} - i{\kappa _3}{A_1}{A_2}A_4^*{{\rm{e}}^{i\Delta kx}} - i{A_3}\sum\limits_{m = 1}^4 {{\alpha _{3m}}{A_m}A_m^*}  = 0 \label{eq:a4:4}\\
\frac{{d{A_4}}}{{dx}} - i{\kappa _4}{A_1}{A_2}A_3^*{{\rm{e}}^{i\Delta kx}} - i{A_4}\sum\limits_{m = 1}^4 {{\alpha _{4m}}{A_m}A_m^*}  = 0 \label{eq:a4:5}
\end{align}
where $\Delta {k_L} = {k_1} + {k_2} - {k_3} - {k_4}$ is the phase mismatch in the weak field limit and the couplings are defined by:
\begin{align}
{\kappa _n} = \frac{{{a^4}{k_1}{k_2}{k_3}{k_4}({k_n} - {\varepsilon _n}\Delta {k_L})}}{{8{C_0}I_j^2{L^3}\omega _n^2}}\\
{\alpha _{nm}} = \frac{{{a^4}k_n^3k_m^2(2 - {\delta _{nm}})}}{{16{C_0}I_j^2{L^3}\omega _n^2}}
\end{align}
where the coupling constant ${\kappa _n}$  describes the four wave mixing process, ${\alpha _{nm}}$    describes self and cross phase modulation, ${\varepsilon _1} = {\varepsilon _2} = 1$ and ${\varepsilon_3} = {\varepsilon_4} =  - 1$ are constants, and $\delta_{nm}$ is the Kronecker delta.
 We solve these coupled complex nonlinear differential equations by converting the complex amplitudes to real amplitudes and finding a solution in the form of elliptic integrals. 
\end{document}